\documentstyle[aps,prl,multicol,epsf]{revtex}

\begin{document}

\title{Competition between the Mott transition and 
the Anderson localization
in 1D disordered interacting electron systems}

\author{Satoshi Fujimoto}
\address{Department of Physics, Kyoto University, Kyoto 606, Japan}

\author{Norio Kawakami}
\address{Department of Applied Physics,
and Department of Material and Life Science, \\
Osaka University, Suita, Osaka 565, Japan}

\date{\today}

\maketitle
     
\begin{abstract}
The competition between the Mott transition and the Anderson 
localization in one dimensional electron systems is studied 
based upon 
the bosonization and the renormalization group method.
The beta function is calculated up to the second order 
in the strength
of diagonal disorder by using a replica trick.
It is found that the sufficiently strong forward scattering 
by random impurities destroys the Mott-Hubbard gap, and  
the backward scattering gives rise to the Anderson 
localization for the resulting gapless state. On the other hand,
if the  Umklapp interaction is strong enough,
the Mott insulating state still overwhelms 
the Anderson localization.
\end{abstract}

\pacs{PACS number: 71.27.+a, 71.30.+h}
\begin{multicols}{2}
There has been continuous interest in the studies
of correlated electron systems in the presence of disorder
\cite{fin,aa,fuku,loc}.
Although in the case of weak localization the effect of 
electron-electron interaction has been extensively studied 
in the Hartree-Fock approximation, it is not well 
understood how the strongly correlated electron systems
such as the Hubbard model are affected by the presence 
of disorder\cite{sand,kot}. In particular, the effect of 
disorder on electron systems with the excitation gap,
such as the Mott insulator,  has received 
considerable attention recently. 

In one-dimensional(1D) electron systems, 
some elaborate techniques like bosonization 
and renormalization group
were successfully applied to disordered systems and 
 revealed the role of electron-electron interaction 
for the localization-delocalization 
transition\cite{chui,appel,suzu,gia}.
In the previous studies, however, the Umklapp interaction which 
is essential to cause the Mott transition was not 
taken into account. To investigate the disorder effect
on the Mott insulating phase, it is crucial to 
study how  the Umklapp interaction should affect 
low-energy properties.

Motivated by this, we investigate 1D 
disordered interacting electron 
systems putting emphasis on the competition between 
the Mott transition and the Anderson localization. 
We clarify the interplay between 
the disorder and the Umklapp interaction 
by using the bosonization and the renormalization group method. 
It is found that the Mott-Hubbard gap collapses 
and a gapless metallic state appears when 
the random forward scattering due to impurities is strong enough. 
The impurity backward scattering  
then drives this metallic state to the Anderson localized state.
We should mention here a recent numerical work 
for the 1D Hubbard model 
with disorder, which indicates the transition from 
the Mott insulator to the Anderson 
localization\cite{sand}.  We note that our field-theoretic 
approach not only provides the knowledge complementary to the 
above numerical work, but also reveals a new feature 
in the role played by the forward and backward 
impurity scatterings. 

We start by introducing the effective Hamiltonian 
to describe 1D interacting electron systems 
with random potentials. Applying standard bosonization rules, 
we write down  the effective Hamiltonian 
in field theory limit\cite{hal},
\begin{eqnarray} 
H&=&H_c+H_s+H_{dis}  \\
H_c&=&\int dx\biggl
[\frac{v_{c}}{2 K_c}
(\partial_x \phi_{c}(x))^2+\frac{v_{c}
K_{c}}{2}(\Pi_{c}(x))^2\biggr] \nonumber \\
&&+ \frac{U}{\alpha^2}
\int dx \cos(\sqrt{8\pi}\phi_c(x)+\delta x), \label{eqn:cham}\\
H_s&=&\int dx\frac{2\pi v_s}{3}[\bbox{J}_{L}(x)\cdot\bbox{J}_{L}(x)+
\bbox{J}_{R}(x)\cdot\bbox{J}_{R}(x)] \nonumber \\
&&+\lambda \int dx 
\bbox{J}_{L}(x)\cdot\bbox{J}_{R}(x), \label{eqn:sham} \\
H_{dis}&=&\sqrt{\frac{2}{\pi}}\int dx 
\eta (x)\partial_x \phi_c(x) \nonumber \\
&+&\frac{1}{\alpha}
\int dx \{\xi(x)e^{i(\sqrt{2\pi}\phi_c(x)+2k_F x)}
{\rm tr}(g(x))+h.c.\}. \label{eqn:dham}
\end{eqnarray}
Here $\phi_c$ is a boson phase field 
for the charge degrees of freedom 
and $\Pi_c$ is its canonical conjugate momentum field, 
and $\delta\equiv 4k_F-2\pi$.
The $U$-term in eq.(\ref{eqn:cham}) 
represents the Umklapp interaction,
which may cause the Mott transition.
For the spin degrees of freedom, we have used non-abelian 
bosonization\cite{witten,aff1} to preserve SU(2) symmetry:
$\bbox{J}_{L(R)}$ is the left(right)-going current 
operator of level-1 SU(2) Kac-Moody algebra, and
$g(x)$ is a fundamental representation of SU(2) Lie algebra.
The $\lambda$-term in eq.(\ref{eqn:sham}) is a marginally irrelevant
interaction which arises from SU(2)$\times$SU(2) symmetry.
We have introduced real and complex random fields 
$\eta(x)$ and $\xi(x)$ for forward and backward 
scatterings by impurities, respectively, which obey the gaussian 
distribution law, 
$\langle \eta(x)\eta (x')\rangle=D_{\eta}\delta(x-x')$,
$\langle \xi^{*}(x)\xi (x')\rangle=D_{\xi}\delta(x-x')$, and 
$\langle \eta(x)\rangle=\langle\xi(x)\rangle=0$.
In addition to the above interactions, 
we should consider the following term
\begin{equation}
H_{rg}= -\sqrt{\frac{2}{\pi}}\int dx A(x)
\partial_x \theta_c, \label{eqn:rang}
\end{equation}
where $\partial_x \theta_c \equiv \Pi_c$ and $A(x)$ 
is a random gauge 
field with the gaussian distribution,
$\langle A(x)A(x')\rangle=D_{A}\delta(x-x')$, 
$\langle A(x)\rangle=0$.
As will be seen later, this term is generated 
in the process of renormalization due to the backward scattering
by random potentials. 
We note that apart from the $U$-term, the charge part of 
the Hamiltonian is similar to that of 
the 2D random phase sine-Gordon model
\cite{hou,cardy} for which randomness 
is introduced for the 2D plane in contrast to 
the present 1D system.

We now consider the quenched disorder with the use of a replica
trick.  Introducing $n$ species of replicas and integrating over
the random variables\cite{gia}, we have the following effective 
action arising from disorder,
\begin{eqnarray}
S_{dis}&=&-\frac{2D_{\eta}}{\pi}\int dx \int d\tau \int d\tau'
\sum_{i,j}\partial_x\phi^i_c(x, \tau)
\partial_x\phi^j_c(x, \tau') \nonumber  \\
&-&\frac{2D_{A}}{\pi}\int dx \int d\tau \int d\tau'
\sum_{i,j}\frac{1}{v_c^2}\partial_{\tau}\phi^i_c(x, \tau)
\partial_{\tau}\phi^j_c(x, \tau') \nonumber \\
&-&\frac{D_{\xi}}{\alpha^2}\int dx \int d\tau \int d\tau'
\sum_{i,j}{\rm tr}(g^i(x, \tau))
{\rm tr}(g^j(x, \tau')) \nonumber \\
&&\times\cos\sqrt{2\pi}(\phi^i_c(x,\tau)-\phi^j_c(x,\tau')),
\end{eqnarray} 
where $i$, $j$ are replica indices.
In the following argument, we calculate the beta functions
up to the second order in $D_{\xi,\eta,A}$.  To this end,
it is quite useful to exploit
the operator product expansions for the U(1) gaussian
model\cite{kada},
\begin{eqnarray}
e^{i\alpha \phi(x,\tau)}e^{-i\alpha \phi(0,0)}&\sim&
\frac{i\alpha}{\vert z\vert^{\alpha^2K_c/2\pi-2}}
(\frac{\partial_z \phi(0,0)}{\bar{z}}
+\frac{\partial_{\bar{z}}\phi(0,0)}
{z}) \nonumber \\
+\frac{1}{\vert z\vert^{\alpha^2K_c/2\pi}} 
&-&\frac{\alpha^2}{\vert z\vert^{\alpha^2K_c/2\pi-2}}
\partial_z\phi(0,0)
\partial_{\bar{z}}\phi(0,0 ) \nonumber \\
-\frac{\alpha^2}{2\vert z\vert^{\alpha^2K_c/2\pi}}
&(&z^2(\partial_z\phi)^2+\bar{z}^2(\partial_{\bar{z}}\phi)^2)
+\cdot\cdot\cdot ,\\
\partial_{z}\phi(x,\tau)e^{i\alpha\phi(0,0)}&\sim&
\frac{i\alpha K_c}{8\pi z}e^{i\alpha\phi(0,0)}+\cdot\cdot\cdot,
\end{eqnarray}
with $z=x+iv_c\tau$, $\bar{z}=x-iv_c\tau$,
 and also those for the level-1 SU(2) Wess-Zumino-Witten model
\cite{zamo,bou,pas},
\begin{eqnarray}
J^a_L(x,\tau)J^b_L(0,0)&\sim&\frac{\delta_{ab}}{z^2}
+\frac{\varepsilon^{abc}}
{z}J^c_L(0,0)+\cdot\cdot\cdot,  \\
J^a_L(x,\tau)g(0,0)&\sim&\frac{t^a}{z}g(0,0)+\cdot\cdot\cdot, \\
{\rm tr}(g(x,\tau)){\rm tr}(g(0,0))&\sim&
\vert z\vert \bbox{J}_{L}(0,0)\cdot\bbox{J}_{R}(0,0)
+\frac{1}{\vert z\vert}  \nonumber \\
&+&\frac{1}{\vert z\vert}
(z^2\bbox{J}_L(0,0)\cdot\bbox{J}_L(0,0) \nonumber \\
&+&\bar{z}^2\bbox{J}_R(0,0)\cdot\bbox{J}_R(0,0))+\cdot\cdot\cdot,
\end{eqnarray}
with $z=x+iv_s\tau$, $\bar{z}=x-iv_s\tau$, and $t^a$, 
the generator of the SU(2) Lie algebra. 
By expanding the action in terms of the interactions and using the 
above operator product expansions, we
 end up with the following scaling equations 
up to the second order in the strength of randomness and the lowest
order in $U$ and $\lambda$ after
taking the replica limit $n\rightarrow 0$, 
\begin{eqnarray}
\frac{d \tilde{D}_{\xi}}{d l}&=&(2-K_c
-3\tilde{\lambda}) \tilde{D}_{\xi}, \label{eqn:sca1}\\
\frac{d \tilde{D}_{\eta}}{d l}&=&\tilde{D}_{\eta}
+4\pi^2 g(u)\tilde{D}_{\xi}^2, \label{eqn:sca2}\\
\frac{d \tilde{D}_{A}}{d l}&=&\tilde{D}_{A}
+4\pi^2 g(u)\tilde{D}_{\xi}^2, \label{eqn:sca3}\\
\frac{d \tilde{U}}{d l}&=&(2-2K_c)\tilde{U}
-\frac{4\tilde{D}_{\eta}K_c^2}{\pi^2}\tilde{U}
, \label{eqn:sca4} \\
\frac{d \tilde{\lambda}}{d l}&=&-\frac{\tilde{\lambda}^2}{2}
-\tilde{D}_{\xi}, 
\label{eqn:sca5}\\
\frac{d K_c}{d l}&=&-2\pi K_c^2\tilde{U}^2J_0(\delta \alpha)
-\frac{K_c^2\tilde{D}_{\xi}}{2u}, \label{eqn:sca6}\\
\frac{d v_c}{d l}&=&-\frac{\pi K_c\tilde{D}_{\xi}v_c}{2u}, 
\label{eqn:sca7}\\
\frac{d u}{d l}&=&-u\tilde{D}_{\xi}, \label{eqn:sca8}
\end{eqnarray}
where $\tilde{D}_{\xi,\eta,A}\equiv D_{\xi,\eta,A}/v_c^2$, 
$\tilde{U}\equiv U/v_c$, $\tilde{\lambda}\equiv \lambda/v_s$,
$u\equiv v_s/v_c$, 
\begin{eqnarray}
g(u)&=&\biggl[\int^{\infty}_{-\infty}
\frac{dy}{\sqrt{1+u^2y^2}(1+y^2)^{K_c/2}}\biggr]^2, 
\end{eqnarray}
and $J_0(x)$ is the Bessel function.
We see from eq.(\ref{eqn:sca3}) that $D_A$-term is 
indeed generated by the second order 
contribution of $D_{\xi}$-term, even when
$D_A$ is initially equal to zero as mentioned before.
We can see that $D_A$  does not couple to the Umklapp term,
so that the random gauge field may not cause any essential
change in the Mott insulating phase. The effect of the random gauge 
field manifests itself in the metallic phase
away from half-filling, as will be discussed later again.
Up to the first order in $D_{\xi}$ and the zeroth order in
$D_{\eta}$ and $D_A$, these scaling equations
coincide with those obtained by Giamarchi and Schulz
except that they did not take into account the Umklapp interaction
\cite{gia,gia2}.

In what follows, we restrict our arguments to the 
case of half-filling. We then put 
$\delta=0$ and $J_0(0)=1$ in eq.(\ref{eqn:sca6}).
In the absence of randomness, the Umklapp interaction 
is relevant, resulting in  the low-energy fixed point 
of the Mott insulator with the charge excitation gap.
Let us first study the effects of the random forward 
scattering $\tilde D_{\eta}$ on this Mott insulating phase.
For a while we neglect the backward scattering, by putting
$\tilde{D}_{\xi}=0$ and $\tilde{D}_A=0$ 
($\tilde{D}_A$-term is generated by
$\tilde{D}_{\xi}$-term).  Then $\lambda$ in 
 (\ref{eqn:sca5}) is scaled to zero 
after the  renormalization, preserving 
spin excitations still massless (the interplay between $\lambda$ and 
$D_{\xi}$  will be discussed later).
Also, from eq.(\ref{eqn:sca2}), we have
$\tilde{D}_{\eta}(l)=\tilde{D}_{\eta}(0)e^{l}$.
Substituting this expression into eq.(\ref{eqn:sca4}), and solving 
eqs.(\ref{eqn:sca4}) and (\ref{eqn:sca6})
numerically, we obtain the renormalization flow for $\tilde{U}$ 
and $\tilde{D}_{\eta}$ as shown in Fig. 1. 
We find that for sufficiently large values of $\tilde{D}_{\eta}(0)$, 
the Umklapp term, $\tilde{U}(l)$,
becomes irrelevant and thus the Mott-Hubbard gap disappears,
resulting in massless charge excitations. 
Note, however, that this massless state is not a conventional 
Tomonaga-Luttinger liquid but 
is a disordered metallic state for which some spatial
correlation functions show exponential decay, 
as discussed in \cite{gia}. On the other hand, 
if $\tilde{U}(0)$ is sufficiently large compared to
 $\tilde{D}_{\eta}(0)$, we can see that
$\tilde{U}(l)$ is scaled to a strong-coupling value
as $l\rightarrow +\infty$, 
and thus the Mott-Hubbard gap should still persist.
The transition between the Mott insulator and the gapless 
metallic phase is of  Kosterlitz-Thouless type\cite{kt}.
Also, the numerical results indicate that at the critical point
the magnitude of the Mott-Hubbard 
gap is roughly proportional to the inverse 
of the correlation length, 
$ 1/\ln(\tilde{D}_{\eta}^{-1}(0))$.

Having noticed that the strong forward scattering can
drive the Mott insulator to a disordered metallic state,
let us now discuss the effects of the backward 
scattering $\tilde D_{\xi}$ which may bring about the Anderson 
localization.
The scaling equation (\ref{eqn:sca1}) implies 
that  for the repulsive $U$ ($K_c<1$), 
$\tilde{D}_\xi$ is renormalized to a
larger value even if its initial value is small, 
and hence the backward scattering becomes relevant. 
Therefore, if there is no intermediate fixed point
 between the weak-coupling and
strong-coupling regimes, the low-energy fixed point is classified 
by $\tilde{U}^{*}=0$ and $\tilde{D}_{\xi}^{*}\rightarrow +\infty$,
or by $\tilde{U}^{*}\rightarrow \infty$ and 
$\tilde{D}_{\xi}^{*}\rightarrow +\infty$.
In both cases, the fixed-point is identified with  
the insulator, since the Drude weight $D=v_cK_c$
is scaled to zero by $\tilde{U}$ or $\tilde{D}_{\xi}$.
In order to see whether this is the Mott insulating state
or the Anderson localized state, 
we examine the behavior of the compressibility.
The compressibility is given by $K_c/v_c$, which satisfies,
\begin{equation}
\frac{d}{dl}\biggl(\frac{K_c}{v_c}\biggr)=-2\pi \tilde{U}^2K_c/v_c,
\label{eqn:comp}
\end{equation}
as easily seen from eqs.(\ref{eqn:sca6}) and (\ref{eqn:sca7}).
Thus if $\tilde{U}$ is irrelevant, the compressibility 
takes a finite value, 
resulting in the fixed point of the Anderson localization. 
On the other hand, if $\tilde{U}$ is relevant, the compressibility 
vanishes at the low-energy fixed point,  
characterizing the Mott insulator.
Therefore when the random forward scattering $\tilde{D}_{\eta}(0)$
is in the region where $\tilde{U}\rightarrow 0$ in Fig. 1, 
the backward scattering $\tilde{D}_{\xi}(0)$ 
drives the system to the Anderson localized state, whereas
for sufficiently large values of $\tilde{U}(0)$ compared to 
$\tilde{D}_{\eta}(0)$, the Mott insulating state still persists. 
Therefore, we end up with the conclusion that 
there can occur the transition driven by disorder
from the Mott insulator to the Anderson insulator
for 1D disordered electron systems, 
although the Mott insulator still overwhelms 
the Anderson localization
if the Umklapp interaction $\tilde{U}(0)$ is large enough.
This is consistent with the numerical results for  
the disordered Hubbard model \cite{sand}.

The above characteristic behavior 
found for correlated electron systems 
shows  sharp contrast to the results for 1D spinless 
fermion systems \cite{shan}. A remarkable point 
for the spinless fermion model is that
the Mott transition  is always accompanied by 
the order of $2k_F$-charge density wave (CDW). Since this 
state can be mapped to the Ising ordered state, 
one can apply Imry-Ma's statement \cite{imry}
that infinitesimally small disorder destroys this Ising 
ordered state.
This indeed leads to the conclusion  
that the Mott insulator for the spinless fermion 
model is changed to the Anderson 
localized state even if infinitesimally small
disorder is introduced \cite{shan}.
Obviously, this argument cannot be directly 
applied to the Mott insulator 
for the present electron systems, which is not accompanied 
by the CDW order.

The difference between  the spinless model and the electron model 
can also be seen in the scaling 
equations (\ref{eqn:sca2}) and (\ref{eqn:sca3}):
for the spinless model $g(u)$ is replaced by 
$(\Gamma(K_c-1/2)/\Gamma(K_c))^2$, 
where $\Gamma(x)$ is the gamma function.
While  $g(u)$ is always finite for 
the electron model,  it diverges for $K_c\leq 1/2$
for the spinless fermion model, and the renormalization group
procedure breaks down. 
Thus as  $K_c$ approaches this singular point, 
$\tilde{D}_{\eta}$ becomes large, and at last 
the right-hand side of eq.(\ref{eqn:sca4}) becomes negative,
even for   infinitesimally small 
$\tilde{D}_{\xi, \eta}(0)$.
Note that   the Mott transition
for the pure spinless model occurs at $K_c=1/2$.
As a result, the Umklapp term $\tilde{U}$ becomes always irrelevant
and the Mott gap closes for a disordered spinless model. 
The above comparison with the spinless 
fermion model naturally leads us
to claim that the stability of the Mott insulator 
for electron systems 
against weak diagonal  disorder may be closely related to 
the absence of the long-range  order.

So far, we have been  concerned with the charge degrees of freedom.
Here we briefly mention the effect of disorder on 
the spin degrees of freedom.
The scaling equation (\ref{eqn:sca5}) indicates that 
the backward scattering  $\tilde{D}_{\xi}$ may
renormalize $\tilde{\lambda}$ to a negative large value, hence
freezing the spin degrees of freedom.
This is essentially the same as that observed for the case 
away from half filling \cite{gia}. Therefore,
the Mott-insulating phase with frozen spins may be realized 
if the spin degrees of freedom is frozen
 after the charge-gap formation due to the Umklapp term
(the large-$U$ region in Fig.1).
Also, for the small-$U$ region,
the Anderson localized state with frozen spins may be expected.

Finally some comments are in order for the random gauge field 
$A(x)$ generated by the impurity backward scattering.
Although  this term does not cause a serious change 
in the half-filled Mott insulator,
 it plays an important role in a delocalized phase.
To see this clearly,  we consider the system away from half-filling. 
We can then put $\tilde{U}=0$ in the scaling equations
because the Umklapp interaction is irrelevant.
We see from eq.(\ref{eqn:sca1}) that if the 
electron-electron interaction 
is strongly attractive and the condition, 
$2-K_c-3\tilde{\lambda}<0$, is satisfied, the backward scattering
$\tilde{D}_{\xi}$ becomes irrelevant and 
then a delocalized metallic phase realizes.
It has been naively expected that in this phase the correlation for
singlet superconductivity is dominant\cite{chui,appel,suzu,gia}.
It turns out, however, the random gauge field may 
change properties of this phase. 
The random gauge field $A(x)$ as well as $\eta(x)$
can be incorporated into the shift of the phase fields\cite{gia}.
Defining new phase fields as,
\begin{eqnarray}
\tilde{\theta}_c(x)&\equiv& \theta_c(x)+\tilde{A}(x),  \\
\tilde{A}(x)&\equiv& \frac{1}{K_cv_c}\sqrt{\frac{2}{\pi}}
\int^x dx' A(x'),
\end{eqnarray}
etc., we can cast the charge part of the Hamiltonian 
into a gaussian type.
Although the random field, $A(x)$, is now eliminated from
the Hamiltonian, it may change long-range behaviors 
of some correlation functions. 
For example, the random gauge field $A(x)$ brings about
the exponential decay of the correlation function for 
singlet superconducting pairing:
\begin{equation}
\langle c_{\uparrow L}(x)c_{\downarrow R}(x)
c^{\dagger}_{\downarrow R}(0)c^{\dagger}_{\uparrow L}(0)\rangle
\sim \frac{1}{\vert x\vert^{1+\frac{1}{K_c}}}
{\rm ln}^{-3/2}\vert x\vert 
e^{-\frac{4D_A\vert x\vert}{K_c^2v_c^2}}. \label{eqn:super}
\end{equation}
This is due to the random phase shift caused by 
the scattering due to random gauge fields. 
Eq.(\ref{eqn:super}) implies that
the susceptibility for singlet superconductivity does not diverge 
in the thermodynamic limit,
$\chi_{ss}(k=0,\omega)\sim \omega^{1/K_c}$.
Thus the fluctuation toward superconductivity is much suppressed
by  random gauge fields. 

In conclusion, 
the transition between the Mott insulator and Anderson insulator
occurs according to the strength of disorder and 
the Umklapp  interaction.
We have found that the random forward scattering by impurities
plays a central role to change the Mott insulating state to a
metallic state, whereas 
the backward scattering drives  the resulting metallic 
state to the Anderson localized state.

This work was partly supported by a Grant-in-Aid from the Ministry
of Education, Science and Culture.

\begin{figure}
\centerline{\epsfxsize=7.5cm \epsfbox{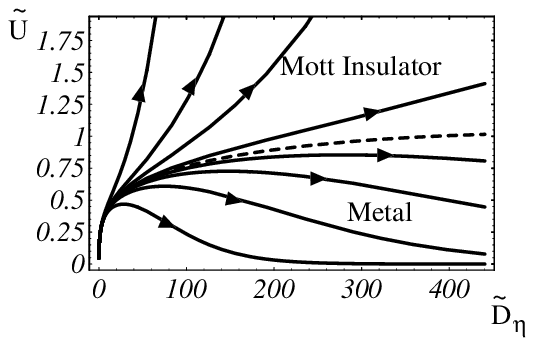}}
{FIG 1. The renormalization flow 
for $\tilde{U}$ and $\tilde{D}_{\eta}$.
The flow for some initial values of $\tilde{U}$ is shown:
$K_c(0)=0.8$, $\tilde{D}_{\eta}(0)=0.05$. 
The transition point exists around $\tilde{U}(0)=0.0463$.}
\end{figure}

\end{multicols}

\end{document}